\documentclass[aps,prl,twocolumn,superscriptaddress,noshowpacs,nofootinbib,amssymb,balancelastpage,preprintnumbers]{revtex4}
\usepackage{graphicx}% Include figure files
\usepackage{wrapfig}
\usepackage{amssymb}
\usepackage{amsmath}
\usepackage{color}
\usepackage{ulem}
\usepackage[breaklinks=true]{hyperref}%Don't forget to erase "pagebackref=true"!!
\begin{document}
\renewcommand{\thefootnote}{\#\arabic{footnote}} 
\setcounter{footnote}{0}
%\clearpage
%%%%%%%%%%%%%%%%%%%%%%%%%%%%%%%%%%%%%%%%%%%%%%%%%%%%%%%%%%%%%%%%%%%%
%%%%% ** Text ** %%%%%%%%%%%%%%%%%%%%%%%%%%%%%%%%%%%%%%%%%%%%%%%%%%%
%%%%%%%%%%%%%%%%%%%%%%%%%%%%%%%%%%%%%%%%%%%%%%%%%%%%%%%%%%%%%%%%%%%%
%
%%%%%%%%%%%%%%%%%%%%%%%%%%%%%%%%%%%%%%%%%%%%%%%%%%%%%%%%%%%%%%%%%%%%%%%
%%%%%  Title Page  %%%%%%%%%%%%%%%%%%%%%%%%%%%%%%%%%%%%%%%%%%%%%%%%%%%%
%\begin{titlepage}
%\begin{center}

%%%%%%%% Preprint #
%\begin{flushright}
%DESY 01-142\\
%\end{flushright}
\preprint{KEK-TH-2292}

%\vskip 2cm

%%%%%%%% Title
\title{
Neutrinoless double beta decays tell nature of right-handed neutrinos
}

%\vskip 1.2cm

%%%%%%%% Authors
\author{Takehiko Asaka}\thanks{{\tt asaka@muse.sc.niigata-u.ac.jp}}
	\affiliation{Department of Physics, Niigata University, Niigata 950-2181, Japan}
\author{Hiroyuki Ishida}\thanks{{\tt ishidah@post.kek.jp}}
	\affiliation{KEK Theory Center, IPNS, Tsukuba, Ibaraki 305-0801, Japan}
\author{Kazuki Tanaka}\thanks{{\tt tanaka@muse.sc.niigata-u.ac.jp}}
	\affiliation{Graduate School of Science and Technology, Niigata University Niigata, 950-2181, Japan}

%\vskip 0.4cm

%%%%%%%% Date
%(\today)
%(April 30, 2010)

%\vskip 2cm
%%%%%%%%%%%%%%%% Abstract %%%%%%%%%%%%%%%%%%%%%%%%%%%%%%%%%%%%%%%%%%%%%
\begin{abstract}
We consider the minimal seesaw model, the Standard Model extended by
two right-handed neutrinos, for explaining the neutrino masses and mixing angles 
measured in oscillation experiments.  
When one of right-handed neutrinos is lighter than 
the electroweak scale, it can give a sizable contribution to neutrinoless
double beta ($0\nu \beta \beta$) decay.
We show that the detection of the $0 \nu \beta \beta$ decay
by future experiments gives a significant implication
to the search for such light right-handed neutrino.
\end{abstract}
%%%%%%%%%%%%%%%% Abstract %%%%%%%%%%%%%%%%%%%%%%%%%%%%%%%%%%%%%%%%%%%%%
%%%%%%%%%%%%%%%%%%%%%%%%%%%%%%%%%%%%%%%%%%%%%%%%%%%%%%%%%%%%%%%%%%%%%%%
%\end{center}
\maketitle
%\end{titlepage}
%%%%%  Title Page  %%%%%%%%%%%%%%%%%%%%%%%%%%%%%%%%%%%%%%%%%%%%%%%%%%%%
%%%%%%%%%%%%%%%%%%%%%%%%%%%%%%%%%%%%%%%%%%%%%%%%%%%%%%%%%%%%%%%%%%%%%%%
%\tableofcontents

%\section{Introduction}

The Standard Model (SM) of the particle physics preserve two accidental global symmetries
in the (classical) Lagrangian, namely 
the baryon and lepton number symmetries. 
It is well known that these global symmetries are non-perturbatively broken 
at the quantum level~\cite{tHooft:1976rip,tHooft:1976snw}, 
especially at high temperature of the universe~\cite{Dimopoulos:1978kv,Manton:1983nd,Klinkhamer:1984di,Kuzmin:1985mm}. 
Even at the quantum level, however, 
a baryon minus lepton symmetry, often called $U(1)_{\rm B \mathchar`- L}$~%
\footnote{We do not specifically consider the symmetry as the gauge symmetry.}, has to be preserved in the SM.

The simplest way to break the $U(1)_{\rm B \mathchar`- L}$ symmetry 
without loss of the renormalizability is 
introducing right-handed neutrinos (RH$\nu$s) into the SM. 
Since RH$\nu$s are singlet under the SM gauge symmetries, 
we can write the mass term, called Majorana mass term, of it without conflicting the gauge principle. 
The Majorana mass term breaks the lepton number symmetry by two units. 
Therefore, the phenomena of the lepton number violation can be 
a definite signal of the existence of RH$\nu$s.

The existence of RH$\nu$s is not only for the violation of the $U(1)_{\rm B \mathchar`- L}$ symmetry 
but also important to solve the origin of the observed tiny neutrino masses. 
In the renormalizable Lagrangian with RH$\nu$s, 
we can obtain two kind of the neutrino masses, 
one is called Dirac masses and another is called Majorana masses. 
When enough hierarchy between these masses is realized, 
we can simply explain the tiny neutrino masses by the seesaw mechanism~%
\cite{Minkowski:1977sc,Yanagida:1979as,Yanagida:1980xy,Ramond:1979,GellMann:1980vs,Glashow:1979,Mohapatra:1979ia}.
In addition, the violation of $U(1)_{\rm B \mathchar`- L}$ can seeds 
the origin of the baryon asymmetry of the universe~%
\footnote{There are a bunch of possibilities to provide 
the baryon asymmetry through the lepton number violation. 
But the detail of the mechanism is independent of the discussions below. 
}. 

One of the most promising signals of the $U(1)_{\rm B \mathchar`- L}$ violation 
is the neutrinoless double beta decay, 
which breaks the lepton number by two units while keeping the baryon number.
(See, for example, articles~\cite{Doi:1985dx,Pas:2015eia,DellOro:2016tmg,Dolinski:2019nrj}.) 
The rate of the decay is characterized by the effective mass 
defined by the neutrino masses and mixing angles. 
When we simply add Majorana masses of three (active or left-handed) neutrinos 
which are responsible for the neutrino oscillation into the SM, 
the effective mass can be predicted depending on 
the lightest active neutrino mass together with the unknown CP violating phases. 
%{\color{red} The predicted range of the effective mass is
%$m_{\rm eff} \le **$~meV for the normal hierarchy (NH) case
%and $m_{\rm eff} =(** - **)$~meV for the inverted (IH) case,
%respectively.
%Here we have assumed that the upper bound
%on the sum of neutrino masses as
%$\sum_i m_i \le **$~eV,
%which is imposed from the cosmological observations.
%}

In view of the fundamental models for the origin of the neutrino masses, 
the mass of the lightest active neutrino cannot be determined uniquely,
leading to different predictions on the effective mass. 
It should be noted that the effective mass can vanish 
in the normal hierarchy (NH) case of the active neutrinos in a certain parameter region.
In such a case, the contribution 
from new physics (other than active neutrinos) including RH$\nu$s would be more 
important for the detection. 
So far, no neutrinoless double beta decay is detected 
and the upper bounds on the effective mass 
have been imposed by various experiments.%
\footnote{In a recent analysis~\cite{1833580}, 
the differential rate of two neutrinoless double beta decay is discussed 
to constrain mixing elements of RH$\nu$s with masses at $\mathcal{O} (0.1 \mathchar`- 10)~{\rm MeV}$.}
The most stringent bound at present is
$61$-$165$~meV by the KamLAND-Zen experiment~\cite{KamLAND-Zen:2016pfg}.
Since this limit is approaching to the predicted range 
in the inverted hierarchy (IH) case, 
the experimental results in near future 
can give us some implication on RH$\nu$s. 

There are several interesting possibilities that the effective mass can be significantly modified 
due to the destructive or constructive contribution from RH$\nu$s.
This additional contribution becomes important when 
the masses of RH$\nu$s are smaller or comparable to the typical scale of Fermi momentum 
in the decaying nucleus ($\sim \mathcal{O}(100)~{\rm MeV}$).

Recently, we have pointed out 
one interesting possibility 
is that RH$\nu$ may hide one of the neutrinoless
double beta decay processes~\cite{Asaka:2020wfo,Asaka:2020lsx} 
(see also Refs.~\cite{Halprin:1983ez,Leung:1984vy}).
This is due to the destructive contribution
of RH$\nu$ to the effective mass.
Note that the impact of RH$\nu$ does 
depend on the decaying nuclei.
If this is the case, 
the mixing elements
of RH$\nu$ can be predicted in terms of its 
mass in a certain range 
which is a good target of future search experiments.

In this paper, we project out the consequences
of the opposite situation, namely
the case when the neutrinoless double beta decay
is observed in some nucleus, and discuss
the impacts on the mixing elements
of RH$\nu$s.

%\section{Minimal Seesaw Model}
First of all, let us explain the framework of the present analysis,
the minimal seesaw model.
It is the Standard Model extended by two right-handed neutrinos
$\nu_{RI}$ ($I=1,2)$, which Lagrangian is given by
%\begin{widetext}
\begin{align}
    {\cal L}
    =&
    {\cal L}_{\rm SM}
    + i \overline {\nu_{RI}}
    \gamma^\mu \partial_\mu \nu_{RI}
    \nonumber \\
    &
    - \left(
    F_{\alpha I} \overline{L_\alpha} \Phi \nu_{RI}
    + \frac{M_I}{2} \overline{\nu_{RI}^c} \nu_{RI}
    + h.c. 
    \right)\,,
\end{align}
%\end{widetext}
where $L_\alpha = ( \nu_{L \alpha}, e_{L \alpha})^T$
($\alpha = e, \mu, \tau)$ and $\Phi$
are the weak doublets of left-handed lepton and Higgs, respectively.
The Yukawa coupling constants and the Majorana masses for neutrinos
are denoted by $F_{\alpha I}$ and $M_I$. 
By assuming that
the Dirac masses $F_{\alpha I}\langle \Phi \rangle$ are much smaller than
the Majorana mass $M_I$, 
the seesaw mechanism works, and the mass eigenstates of neutrinos
are three active neutrinos $\nu_{i}$ ($i=1,2,3)$ with masses $m_i$
and two heavy neutral leptons (HNLs) $N_I$ with masses $M_I$.

The mass ordering of active neutrinos is not determined by the oscillation data,
and two possibilities, the normal hierarchy (NH) with $m_3 > m_2 > m_1=0$
and the inverted hierarchy (IH) with $m_2 > m_1 > m_3=0$, are allowed.
Note that the lightest active neutrino is massless in the considering situation.
On the other hand, we can take the masses of HNLs as $M_2 \ge M_1$ without loss of
generality.  The left-handed (flavor) neutrinos are then written as
\begin{align}
    \nu_{L \alpha}
    = \sum_i U_{\alpha i} \, \nu_i + \sum_I \Theta_{\alpha I} \, N_I^c \,,
\end{align}
where $U_{\alpha i}$ is the mixing matrix of active neutrinos 
called as the PMNS matrix while $\Theta_{\alpha I}$ is that of HNLs.

%\textit{\textbf{$0\nu \beta \beta$ decay and search for HNL: }}~
%\section{%$0\nu \beta \beta$ decay and search for HNL
%Neutrinoless double beta decay and search for HNL}
One of the most important consequences of the seesaw mechanism
is that active neutrinos and HNLs are both Majorana particles.
In this case the lepton number violating processes are induced 
by these particles, which is a clear signature of physics beyond
the SM.  One promising example is the $0 \nu \beta \beta$ decay,
and the quest for the decay is going on by various experiments.

The rate for the $0 \nu \beta \beta$ decay 
mediated by active neutrinos and HNLs is proportional
$|m_{\rm eff}|^2$, where $m_{\rm eff}$ is the so-called 
effective (neutrino) mass in the $0 \nu \beta \beta$ decay.
In the minimal seesaw model it is given by
\begin{align}
    m_{\rm eff} = m_{\rm eff}^\nu + m_{\rm eff}^N \,.
\end{align}
Here the first term in the right-hand side represents the 
contributions from the active neutrinos, which is given by
\begin{align}
    m_{\rm eff}^\nu
    &= \sum_i U_{ei}^2 \, m_i \,.
\end{align}
On the other hand, the contributions from HNLs are expressed as
\begin{align}
    m_{\rm eff}^N
    &= \sum_{I} \Theta_{eI}^2 \, M_I \, f_\beta (M_I) \,, 
\end{align}
where $f_\beta$ is the suppression factor compared to 
$m_{\rm eff}^\nu$ due to the heaviness of HNLs $M_I \gg m_i$.
Here we apply the result in Ref.~\cite{Faessler:2014kka,Barea:2015zfa}
and assume the following form
\begin{align}
    \label{eq:f_beta}
    f_\beta (M) = \frac{\Lambda_\beta^2}
    {\Lambda_\beta^2 + M^2} \,,
\end{align}
where $\Lambda_\beta = {\cal O}(10^2)$~MeV denotes
the typical scale of the Fermi momentum in the $0 \nu \beta \beta$ decay.
Hereafter we take $\Lambda_\beta =200$~MeV as a representative value.

In this letter we consider the impacts of the detection
of the $0 \nu \beta \beta$ decay by future experiments
on the properties of HNLs.
The measurement of the decay rate gives the value
of $|m_{\rm eff}|$.  Note that $m_{\rm eff}$ is a complex number.
First, we consider the case when right-handed neutrinos possess
the hierarchical masses $M_2 \gg M_1$.
We then find that 
the mixing element $|\Theta_{e1}|^2$ of the lighter HNL 
is given by
\begin{align}
    \Theta_{e1}^2 = \frac{m_{\rm eff} - m_{\rm eff}^\nu \left[ 1 - f_\beta (M_2) \right]}
    {M_1 \left[ f_\beta(M_1) - f_\beta(M_2) \right]} \,.
    \label{eq:Eq_THsqe1}
\end{align}
Here we have used the intrinsic relation 
between mixing elements in the seesaw mechanism
\begin{align}
    0 &= \sum_i U_{ei}^2 \, m_i +
    \sum_{I} \Theta_{eI}^2 \, M_I \,.
        \label{eq:SS}
\end{align}
Importantly, the mixing element $|\Theta_{e1}|^2$
is given by $m_{\rm eff}$ and $m_{\rm eff}^\nu$ together with
masses $M_1$ and $M_2$.
This means that, if $|m_{\rm eff}|$ is found by 
the detection of the $0 \nu \beta \beta$ decay,
the range of $|\Theta_{e1}|^2$ can be predicted.
In practice both upper and lower bounds on $|\Theta_{e1}|^2$
are obtained by varying the unknown parameters in $m_{\rm eff}^\nu$
({\it i.e.}, the Majorana phase $\eta$ and the mass ordering) 
and the phase of $m_{\rm eff}$.

%%%%%%%%%%%%%%%%%%%%%%%%%%%%%%%%%%%%%%%%%%%%%%%%%%%%%%%%%%%%%%%%%%%%%%%
%%%%% ** Figure ** %%%%%%%%%%%%%%%%%%%%%%%%%%%%%%%%%%%%%%%%%%%%%%%%%%%%
\begin{figure}[t]
  \centerline{
  \includegraphics[width=4.8cm]{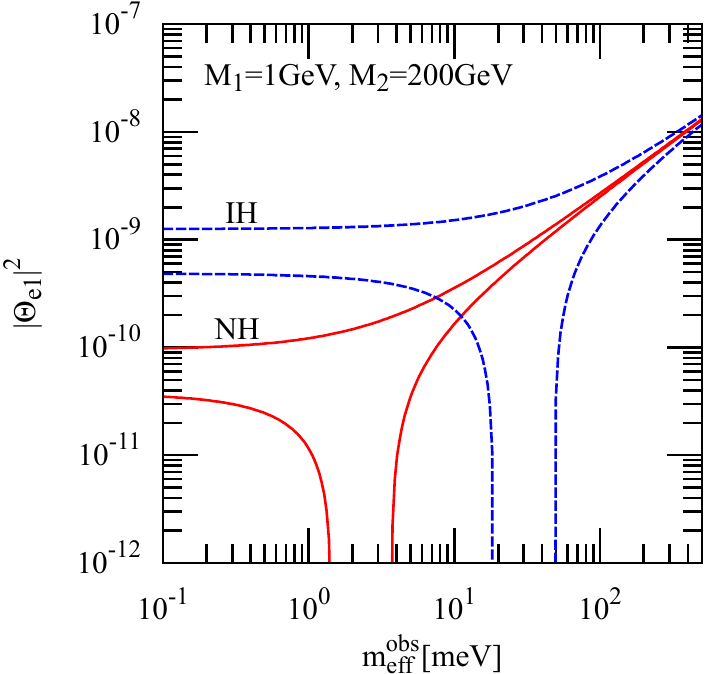}%
  }%
  \vspace{-2ex}
  \caption{
		Upper and lower bounds on $|\Theta_{e1}|^2$ 
		for the NH (red solid lines) and IH (blue dashed lines) cases.	
		Here $M_1=1$~GeV and $M_2 =200$~GeV.
  }
  \label{fig:FIG_THsqe1_M1=1GeV_M2=200GeV}%
  %}%}
\end{figure}
%%%%%%%%%%%%%%%%%%%%%%%%%%%%%%%%%%%%%%%%%%%%%%%%%%%%%%%%%%%%%%%%%%%%%%%
When $M_1=1$~GeV and $M_2=200$~GeV, these bounds are 
shown in Fig.~\ref{fig:FIG_THsqe1_M1=1GeV_M2=200GeV} in
terms of the (would-be) observed value of $|m_{\rm eff}|$
denoted by $m_{\rm eff}^{\rm obs}$.
In the present analysis we take the central values 
of the mass squared differences, the mixing angles and the Dirac phase in the PMNS matrix 
given in Ref.~\cite{Esteban:2020cvm} for the estimation of 
$|m_{\rm eff}^\nu|$.  We find that 
$|m_{\rm eff}^\nu| = 1.45$--$3.68$~meV
and 18.6--48.4~meV for the NH and IH cases, respectively.
It is found from Eq.~(\ref{eq:Eq_THsqe1})
that the lower bound on $|\Theta_{e1}|^2$ vanishes
when $m_{\rm eff}^{\rm obs} = |m_{\rm eff}^\nu|( 1 - f_\beta (M_2))$. 
%{\color{red}\sout{
%Note that, when $m_{\rm eff}^{\rm obs} \ll |m_{\rm eff}^\nu|$, 
%the mixing element is given by}}
%\begin{align}
%    {\color{red}\mbox{[No need?]}
%    |\Theta_{e1}^2| = \frac{|m_{\rm eff}^\nu| \left[ 1 - f_\beta (M_2) \right]}
%    {M_1 \left[ f_\beta(M_1) - f_\beta(M_2) \right]} \,.}
%    \label{eq:Eq_THsqe1-2}
%\end{align}
%{\color{blue} \bf (I'm agree with erasing this equation since we don't show any result anywhere. [HI])}

%%%%%%%%%%%%%%%%%%%%%%%%%%%%%%%%%%%%%%%%%%%%%%%%%%%%%%%%%%%%%%%%%%%%%%%
%%%%% ** Figure ** %%%%%%%%%%%%%%%%%%%%%%%%%%%%%%%%%%%%%%%%%%%%%%%%%%%%
\begin{figure}[t]
  \centerline{
  \includegraphics[width=4.5cm]{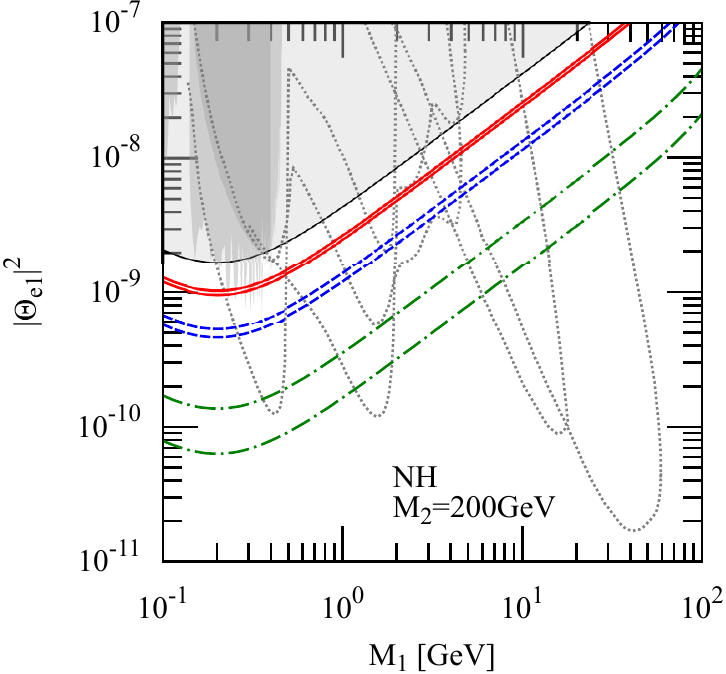}%
  \includegraphics[width=4.5cm]{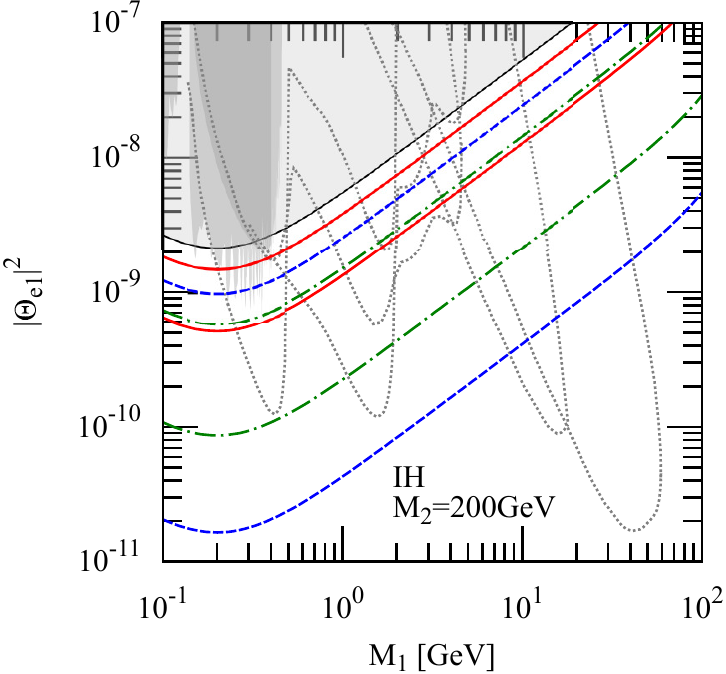}%
  }%
  \vspace{-2ex}
  \caption{
		Upper and lower bounds on $|\Theta_{e1}|^2$ 
		for the NH (left) and IH (right) cases.	
		We take $m_{\rm eff}^{\rm obs}$ =100~meV (red sold lines),
		50~meV (blue dashed lines), and 10~meV (green dot-dashed lines).
		Here $M_2 =200$~GeV.
		The current (conservative) upper bound on $|\Theta_{e1}|^2$ 
		from $|m_{\rm eff}|<165$~meV is shown by black solid line
		(and the light-gray region is exluded).
		The dark-gray regions are excluded by the direct search experiments.
		The dotted lines shows the sensitivities by the future experiments.
        See the detail in the main text. 
  }
  \label{fig:FIG_THsqe1}
\end{figure}
%%%%%%%%%%%%%%%%%%%%%%%%%%%%%%%%%%%%%%%%%%%%%%%%%%%%%%%%%%%%%%%%%%%%%%%
The predicted range of $|\Theta_{e1}|^2$ is shown in Fig.~\ref{fig:FIG_THsqe1} 
where the current upper bounds and the sensitivities on $|\Theta_{e1}|^2$
by future search experiments are also shown~% 
\cite{PIENU:2011aa,Aguilar-Arevalo:2017vlf,NA62:2020mcv,Blondel:2014bra,SHiP:2018xqw,Krasnov:2019kdc,Alpigiani:2020tva}.  
We take the (would-be) observed value of the effective mass as
$|m_{\rm eff}|$ = 100~meV, 50~meV, and 10~meV.
Importantly, the most of the predicted range can be tested by 
the future experiments.

We should note that the understanding of $f_\beta (M)$ is important
for the precise prediction of the mixing elements, since it contains 
the uncertainty of the order unity.
For this purpose
the better understanding of the nuclear matrix elements
of the $0\nu \beta \beta$ decay mediated by HNL is crucial.

Next, let us consider the case when the masses of HNLs are degenerate
\begin{align}
    M_1 = M_2 = M_N \,.
\end{align}
In this case, the total effective mass is given by
\begin{align}
    m_{\rm eff} = m_{\rm eff}^\nu \left[ 1 - f_\beta (M_N) %(\overline{M})
    \right]\,,
\end{align}
and hence the total value is always smaller than the that from active neutrinos 
$|m_{\rm eff}| < | m_{\rm eff}^\nu|$ as long as HNLs participate the $0 \nu \beta \beta$ decay.  
Note that the arguments of $m_{\rm eff}$ and $m_{\rm eff}^\nu$
are the same.
In this case, we find the interesting consequences
if $|m_{\rm eff}|$ is measured:
First, the mass of degenerate HNLs is determined
depending on the measured value of $|m_{\rm eff}|$ as
\begin{align}
    %\overline{M}
    M_N
    = \Lambda_\beta 
    \sqrt{ \frac{m_{\rm eff}^{\rm obs}}{|m_{\rm eff}^\nu| - m_{\rm eff}^{\rm obs}}} \,.
\end{align}
This shows that, once $m_{\rm eff}^{\rm obs}$ is fixed,
the unknown Majorana phase
in $m_{\rm eff}^\nu$ determines $M_N$.
Second, the sum of the mixing elements is found to be
\begin{align}
    \left| \Theta_{e1}^2 + \Theta_{e2}^2 \right|
    = \frac{|m_{\rm eff}^\nu|}{\Lambda_\beta}
    \sqrt{ \frac{|m_{\rm eff}^\nu| - m_{\rm eff}^{\rm obs}}{m_{\rm eff}^{\rm obs}} }\,.\label{Eq:deg-mix}
\end{align}
%%%%%%%%%%%%%%%%%%%%%%%%%%%%%%%%%%%%%%%%%%%%%%%%%%%%%%%%%%%%%%%%%%%%%%%
%%%%% ** Figure ** %%%%%%%%%%%%%%%%%%%%%%%%%%%%%%%%%%%%%%%%%%%%%%%%%%%%
\begin{figure}[t]
  \centerline{
  \includegraphics[height=4.5cm]{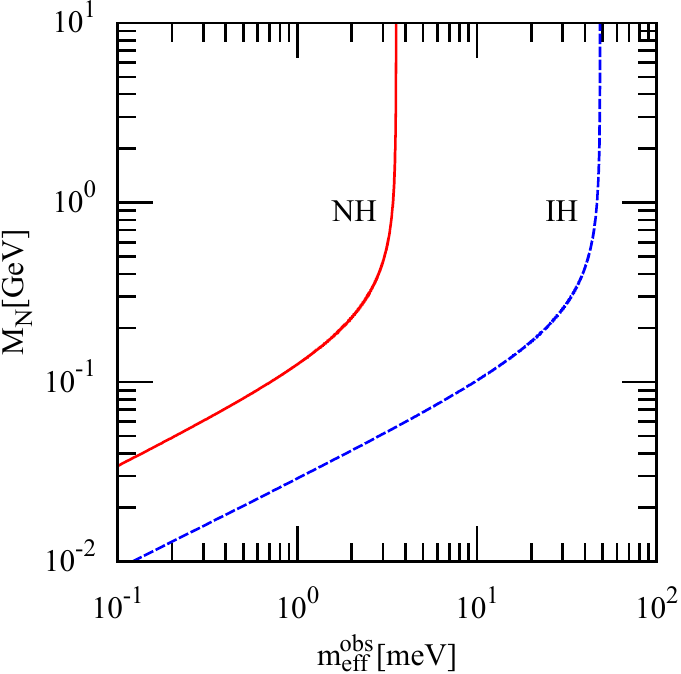}%
  \includegraphics[height=4.5cm]{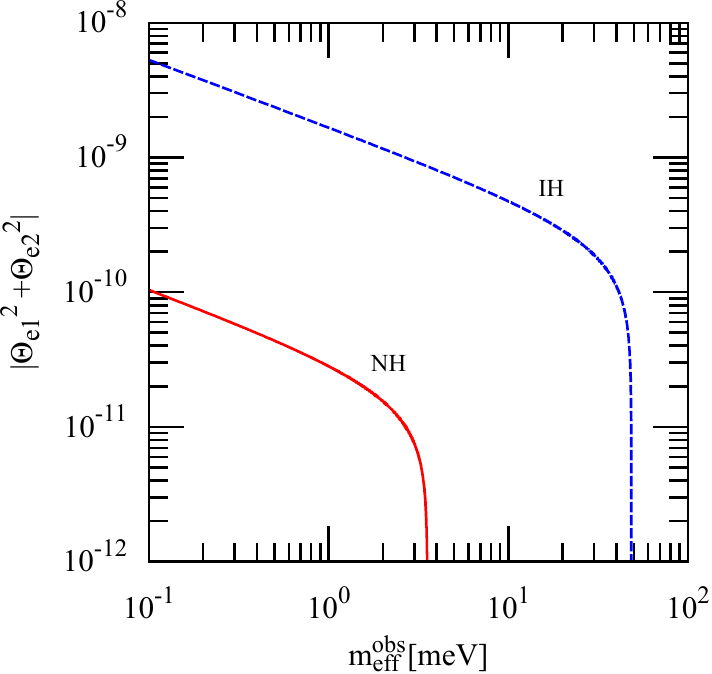}%
  }%
  \vspace{-2ex}
  \caption{
		The degenerate mass $M_N$ and mixing element $|\Theta_{e1}^2+\Theta_{e2}^2|$
		in terms of the observed value $m_{\rm eff}^{\rm obs}$
		in the NH (red solid line) or IH (blue dashed line).
		We take the Majorana phase $\eta = 0$.
  }
  \label{fig:FIG_MN_deg}
\end{figure}
%%%%%%%%%%%%%%%%%%%%%%%%%%%%%%%%%%%%%%%%%%%%%%%%%%%%%%%%%%%%%%%%%%%%%%%
%%%%%%%%%%%%%%%%%%%%%%%%%%%%%%%%%%%%%%%%%%%%%%%%%%%%%%%%%%%%%%%%%%%%%%%
%%%%% ** Figure ** %%%%%%%%%%%%%%%%%%%%%%%%%%%%%%%%%%%%%%%%%%%%%%%%%%%%
\begin{figure}[t]
  \centerline{
  \includegraphics[height=4.5cm]{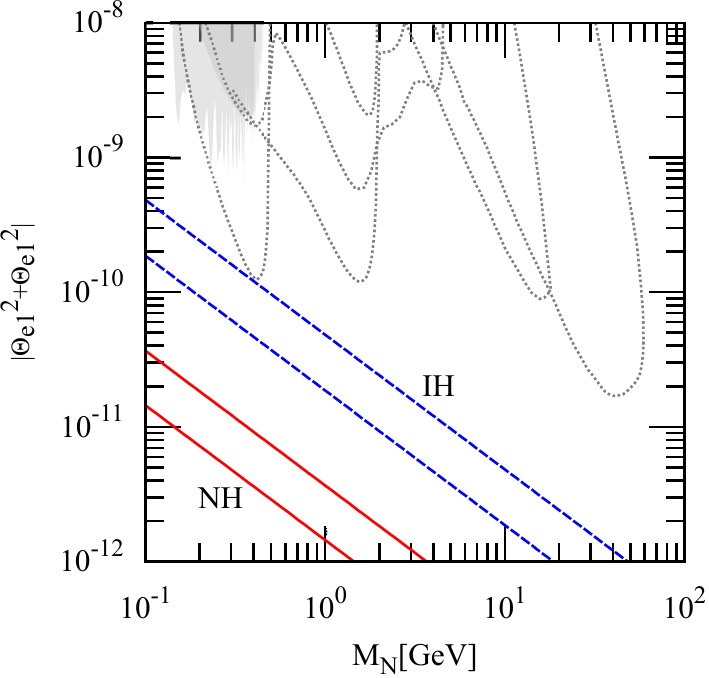}%
  }%
  \vspace{-2ex}
  \caption{
    Range of the mixing element $|\Theta_{e1}^2+\Theta_{e2}^2|$
    in terms of the degenerate mass $M_N$ 
    by taking the Majorana phase $\eta = 0$--$\pi$
    in the NH (red solid line) or IH (blue dashed line).}
  \label{fig:FIG_MN_THsqe1_deg}
\end{figure}
%%%%%%%%%%%%%%%%%%%%%%%%%%%%%%%%%%%%%%%%%%%%%%%%%%%%%%%%%%%%%%%%%%%%%%%

These results are shown in Fig.~\ref{fig:FIG_MN_deg}.
Here we take the Majorana phase as $\eta = 0$,
and $|m_{\rm eff}^\nu|$ = 3.54~meV and 48.4~meV for the NH and IH cases,
respectively.
It is seen that
the observed effective mass 
$m_{\rm eff}^{\rm obs}$ of a few $10$~meV 
corresponds to the Majorana mass $M_N \simeq {\cal O}(0.1-1)$~GeV
and the mass ordering is the IH 
since HNL contributions are always destructive to the active neutrino ones. 
%{\color{red}
%Once we observe the $0 \nu \beta \beta$ decay within 
%the range of the effective mass predicted by the active neutrinos, 
%we can constrain the parameters if degenerated HNLs. }
The relation between $M_N$ and $|\Theta_{e1}^2 + \Theta_{e2}^2|$ is 
shown in Fig.~\ref{fig:FIG_MN_THsqe1_deg}.
%\begin{align}
%    \left| \Theta_{e1}^2+\Theta_{e2}^2 \right| 
%    = 
%    \frac{|m_{\rm eff}^{\nu}|}
%    {M_N} \,,
%\end{align}
%which is shown in Fig.~\ref{fig:FIG_MN_THsqe1_deg}.
We find that in order to test the degenerate case
the improvement of the sensitivity by future experiments is 
required especially for the NH case.

%Notice that we can write the relation between $M_N$ and $|\Theta_{e1}^2 + \Theta_{e2}^2|$ 
%by using $m_{\rm eff}^{\rm obs}$ as 
%\begin{align}
%    \left| \Theta_{e1}^2+\Theta_{e2}^2 \right| 
%    = 
%    \left| 
%    \frac{m_{\rm eff}^{\rm obs}}{\left| f_\beta (M_N) -1 \right| M_N}
%    \right|
%\,.
%\end{align}
%Namely, once $m_{\rm eff}^{\rm obs}$ is fixed, 
%we can predict the mixing elements in terms of the HNL mass
%being independent of the Majorana phase. 

Before concluding the paper, we stress the impact of the difference 
among the $0\nu \beta \beta$ decay nuclei~\cite{Asaka:2020lsx}. 
Throughout this paper, 
we have assumed the approximated form of the suppression function $f_\beta$ 
to be Eq.~(\ref{eq:f_beta}) 
and fixed the typical Fermi momentum as $\Lambda_\beta=200~{\rm MeV}$. 
The important point is that 
the nuclear matrix elements including the suppression factor due to HNLs 
are different depending on the decaying nuclei used in the 
$0\nu \beta \beta$ experiments.
This effect may be quantified by the choice 
the typical Fermi momentum in this analysis. 

%%%%%%%%%%%%%%%%%%%%%%%%%%%%%%%%%%%%%%%%%%%%%%%%%%%%%%%%%%%%%%%%%%%%%%%
%%%%% ** Figure ** %%%%%%%%%%%%%%%%%%%%%%%%%%%%%%%%%%%%%%%%%%%%%%%%%%%%
\begin{figure}[t]
  \centerline{
  \hspace{5mm}
  \mbox{\includegraphics[height=4.4cm]{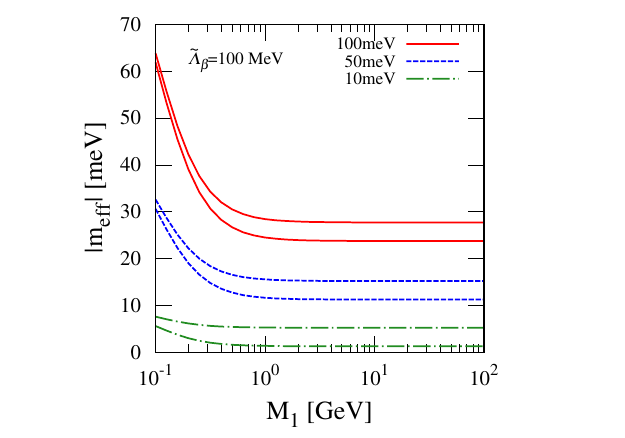}}%
  \hspace{-2cm}
  \includegraphics[height=4.4cm]{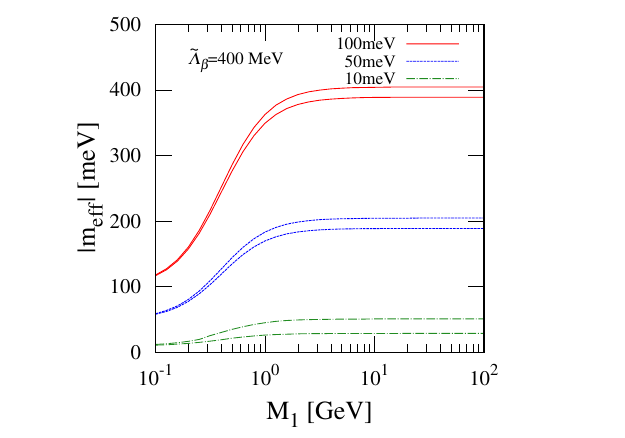}%
  }%
  \vspace{-2ex}
  \caption{
		Upper and lower bounds of predicted effective mass 
		with $\tilde{\Lambda}_\beta = 100~{\rm MeV}$ (left) 
		and $\tilde{\Lambda}_\beta = 400~{\rm MeV}$ (right) in the NH case.
		We assume that the effective mass 
		observed in the nucleus with $\Lambda_\beta = 200~{\rm MeV}$ 
		is $100~{\rm meV}$ (red, solid), $50~{\rm meV}$ (blue, bashed), and $10~{\rm meV}$ (green, dot-dashed).
		Here, we fix $M_2 =200~{\rm GeV}$. 
  }
  \label{fig:FIG_M1_meff_diffLb}
\end{figure}
%%%%%%%%%%%%%%%%%%%%%%%%%%%%%%%%%%%%%%%%%%%%%%%%%%%%%%%%%%%%%%%%%%%%%%%
In Fig.~\ref{fig:FIG_M1_meff_diffLb}, 
we plot the upper and lower values of the predicted effective mass 
with different Fermi momentum from $200~{\rm MeV}$ 
while assuming the $0\nu \beta \beta$ decay is observed at the experiment 
with $\Lambda_\beta=200~{\rm MeV}$ in the NH case. 
We can obtain similar behavior straightforwardly in the IH case as well. 
We take the observed value of the effective mass 
to be $100~{\rm meV}$, $50~{\rm meV}$, or $10~{\rm meV}$. 
Interestingly, the predicted effective mass can be significantly enhanced 
when $\Lambda_\beta$ becomes larger enough than $200~{\rm MeV}$ and $M_1$ gets heavier. 
By inserting Eq.~(\ref{eq:Eq_THsqe1}) into the expression of the effective mass, 
we can obtain 
\begin{align}
    \tilde{m}_{\rm eff} 
    &= 
    \left[ 1 - \tilde{f}_\beta (M_2) \right]
    m_{\rm eff}^\nu \notag\\
    &\hspace{-3mm}
+ 
    \left[
    m_{\rm eff} - m_{\rm eff}^\nu \left[ 1 - f_\beta (M_2) \right]
    \right]
    \frac{\tilde{f}_\beta(M_1) - \tilde{f}_\beta(M_2)}
    {f_\beta(M_1) - f_\beta(M_2)}\,, \label{Eq:meff-diff}
\end{align}
where $\Lambda_\beta = 200$~MeV in $f_\beta$
but $\Lambda_\beta \neq 200$~MeV in $\tilde{f}_\beta$ 
which is denoted as $\tilde{\Lambda}$. 
Since the last fraction in the RHS of Eq.~(\ref{Eq:meff-diff}) can be simplified as 
\begin{align}
\hspace{-1mm}
\frac{\tilde{f}_\beta(M_1) - \tilde{f}_\beta(M_2)}
    {f_\beta(M_1) - f_\beta(M_2)}
    =
        \frac{\tilde{\Lambda}_\beta^2}{\Lambda_\beta^2} 
    \frac{\left( \Lambda_\beta^2 + M_1^2 \right) \left( \Lambda_\beta^2 + M_2^2 \right)}{\left( \tilde{\Lambda}_\beta^2 + M_1^2 \right) \left( \tilde{\Lambda}_\beta^2 + M_2^2 \right)}\,,
\end{align}
Namely, the effective mass is enhanced as $M_1$ gets greater/suppressed than 
the typical Fermi momentum in $\tilde{f}_\beta$ by the factor $\tilde{\Lambda}_\beta^2/\Lambda_\beta^2$.
As clearly seen, since significant enhancement/suppression could happen
depending on the values of $\Lambda_\beta$ due to the contributions from HNLs.
Thus, we can claim that 
the multiple detection by the $0 \nu \beta \beta$ experiments
using different nuclei is crucial to reveal the properties of HNLs.

%\section{Conclusions}
In conclusions, we have considered the minimal seesaw model with two right-handed neutrinos.
It has been shown that, if the effective mass in the $0\nu \beta \beta$ decay 
will be measured by future experiments, the possible range of the mixing elements 
for the lighter heavy neutral lepton (right-handed neutrino) is determined.
Especially, when two heavy neutral leptons are hierarchical and the lighter mass is below 
the electroweak scale, $N_1$ is a good target of the direct search experiments.

It has also been shown that the predicted effective mass 
can depend on nucleus of the experiment. 
Therefore, comprehensive studies on the neutrinoless double beta decays in the seesaw mechanism 
is necessary to extract the concrete information of the heavy neutral leptons.

\subsection*{Acknowledgments}
The work of T.A. was partially supported by JSPS KAKENHI
Grants No. 17K05410, No. 18H03708, No. 19H05097, and No. 20H01898.
The work of H.I. was supported by JSPS KAKENHI Grant No. 18H03708.

\end{document}